\documentclass[12pt]{article}
\usepackage{epsfig}

\title{\Large \bf Axial anomaly in quantum electro- and
 \\  chromodynamics and the structure of the vacuum\\ in quantum chromodynamics}
\author{B.L.Ioffe\\ \\
Institute of Theoretical and Experimental Physics,\\
 B.Cheremushkinskaya 25, 117218,
Moscow,Russia}
\begin{document}
\date{}
\maketitle

\newcommand{\be}{\begin{equation}}
\newcommand{\ee}{\end{equation}}

\def\la{\mathrel{\mathpalette\fun <}}
\def\ga{\mathrel{\mathpalette\fun >}}
\def\fun#1#2{\lower3.6pt\vbox{\baselineskip0pt\lineskip.9pt
\ialign{$\mathsurround=0pt#1\hfil##\hfil$\crcr#2\crcr\sim\crcr}}}

\bigskip

\begin{abstract}

In this  report, I discuss the current state of the problem of the axial anomaly in
quantum electrodynamics (QED) and quantum chromodynamics (QCD) and how the axial anomaly
is related to the structure of the vacuum in QCD. In QCD, the vacuum average of the axial
anomaly is proportional to a new quantum number $n$, the winding number. The axial
anomaly condition implies that there are zero modes of the Dirac equation for a massless
quark and that there is spontaneous breaking of chiral symmetry in QCD, which leads to
the formation of a quark condensate. The axial anomaly can be represented in the form of
a sum rule the structure function in the dispersion representation of the axial -- vector
-- vector (AVV) vertex. On the basis of this sum rule, it is calculated the width of the
$\pi^0\to 2\gamma$ decay with an accuracy of 1.5\%. It is demonstrated, that  $'$t Hooft
conjecture  -- the singularities of the  amplitudes calculated in perturbative QCD on
quark-gluon basis should reproduce themselves in calculations on the hadrons basis -- is
not fulfilled generally.

\end{abstract}

\bigskip

PACS ~numbers: 11.15.-q, 11.30.Qc, 12.38Aw

\vspace{1.5cm}

{\bf \large 1. The definition of an anomaly }

\bigskip

\noindent Let us  suppose that the classical  filed-theory Lagrangian has a certain
symmetry, i.e., is invariant under transformations of the field corresponding to this
symmetry. According to the Noether theorem, the symmetry corresponds to a conservation
law. An anomaly is called a phenomenon  in which the given symmetry and the conservation
law are violated as we pass to quantum theory. The reason for such violation lies in the
singularity of quantum field operators at small distances, such that finding the physical
quantities requires fixing not only the Lagrangian but also the renormalization
procedure. (See reviews dealing with various anomalies in Refs \cite{1}--\cite{4}.)

There ate two types of anomalies, internal and external. In the first case, the gauge
invariance of the classical Lagrangian is broken at the quantum level, the theory becomes
unrenormalizable, and is not self-consistent. This problem can be resolved by a special
choice  of fields in the Lagrangian, for which all the internal anomalies cancel out.
(Such an approach is used in the standard model of electroweak interaction and is known
as the Glashow-Illiopoulos-Maiani mechanism.) External anomalies emerge as a result of
the interaction between the fields in the Lagrangian and external sources. It is these
anomalies that appear in quantum  electrodynamics and quantum chromodynamics; they are
discussed in what follows. We show that anomalies play an important role in QED and
especially  in QCD. Hence, the term ``anomaly'' should not mislead us -- it is a normal
and important ingredient of most quantum field theories.

\vspace{7mm}

{\bf \large 2. Axial anomaly in QED}

\bigskip

\noindent The Dirac equation for an electron in an external electromagnetic field
$A_{\mu}(x)$ is
 \be i\gamma_{\mu} \frac{\partial\psi(x)}{\partial x_{\mu}} =
m\psi(x)-e\gamma_{\mu}A_{\mu}(x)\psi(x).\label{1}\ee The axial current is defined as
\be j_{\mu 5}(x) = \bar{\psi}(x) \gamma_{\mu}\gamma_5\psi(x).\label{2}\ee Its divergence
calculated in classical theory, i.e., with the use of Eqn (\ref{1}), is
 \be
\partial_{\mu}j_{\mu 5}(x) = 2im \bar{\psi}(x) \gamma_5 \psi(x)\label{3}\ee and tends
 to zero as $m \to 0$.  In quantum theory, the axial current must be redefined, because
$j_{\mu 5}(x)$ is the product of two local fermionic fields, with the result that it is
singular when both fields are at the same point.  (Naturally, a similar statement is true
for a vector current.) To achieve a meaningful approach, we split the points where the
two fermionic fields act by a distance $\epsilon$, such that
\be j_{\mu 5} (x,\varepsilon) =\bar{\psi} \biggl ( x +\frac{\varepsilon}{2}\biggr )
\gamma_{\mu}\gamma_5 \exp \biggl [ ie \int\limits^{x+\varepsilon/2}_{x-\varepsilon/2}
dy_{\alpha} A_{\alpha} (y) \biggr ] \psi \biggl (x-\frac{\varepsilon}{2}\biggr
),\label{4}\ee and take  $\varepsilon \to 0$ in the final result. The exponential factor
(\ref{4}) is introduced to ensure the local gauge  invariance of  $j_{\mu 5}
(x,\epsilon)$. The divergence of axial current  (\ref{4}) has the following form (we use
Eqn (\ref{1}) and keep only the terms that are linear in $\epsilon$):
\be
\partial_{\mu}j_{\mu 5} (x,\varepsilon) = 2 im \bar{\psi}
\biggl (x+\frac{\varepsilon}{2}\biggr ) \gamma_5 \psi \biggl(x
-\frac{\varepsilon}{2}\biggr ) - ie\varepsilon_{\alpha} \bar{\psi} \biggl
(x+\frac{\varepsilon}{2}\biggr ) \gamma_{\mu}\gamma_5\psi \biggl (x
-\frac{\varepsilon}{2}\biggr ) F_{\alpha \mu},\label{5}\ee where $F_{\alpha\mu}$ is the
electromagnetic field strength. For simplicity, we assume that   $F_{\mu\nu}=$~const and
use the fixed-point gauge (the Fock-Schwinger gauge)  $x_{\mu} A_{\mu}(x)=0$. Then
$A_{\mu}(x) = (1/2) x_{\nu} F_{\nu \mu}$. We calculate the vacuum average of (\ref{5}).
To calculate the right-hand side of (\ref{5}), we use the expression for the electron
propagator in an external electromagnetic field   $(\not\!x = x_{\mu} \gamma_{\mu})$:
\be S(x) = \frac{i}{2\pi^2}\biggl [\frac{\not\!x}{x^4} +i\frac{m}{2x^2}+\frac{1}{16x^2}
eF_{\mu\nu} (\not\!x\sigma_{\mu\nu} +\sigma_{\mu\nu}\not\!x)\biggr ],\label{6}\ee
\be \sigma_{\mu\nu} =\frac{i}{2} (\gamma_{\mu}\gamma_{\nu} -\gamma_{\nu} \gamma_{\mu}),
\label{7}\ee Vacuum averaging involves first order corrections in $e^2$. Substituting Eqn
(\ref{6}) in Eqn (\ref{5}) and ignoring the electron mass, we obtain
 \be \langle 0 \mid
\partial_{\mu}j_{\mu 5} \mid 0 \rangle =\frac{e^2}{4\pi^2}
F_{\alpha\mu} F_{\lambda \sigma} \varepsilon_{\beta\mu\lambda
\sigma}\frac{\varepsilon_{\alpha}\varepsilon_{\beta}}{\varepsilon^2}. \label{8}\ee
Because there can be no preffered direction in spacetime, the limit $\varepsilon \to 0$
can be achieved in a symmetric manner, and we have
 \be
\partial_{\mu}j_{\mu 5} =\frac{e^2}{8\pi^2} F_{\alpha \beta}
\tilde{F}_{\alpha \beta},\label{9}\ee where
\be \tilde{F}_{\alpha\beta} =\frac{1}{2} \varepsilon_{\alpha \beta \lambda
\sigma}F_{\lambda \sigma}\label{10}\ee is the dial electromagnetic field tensor. In Eqn
(\ref{9}), the symbol of vacuum averaging is dropped because in the  $e^2$-order, this
equation can be considered as an operator equation. Equation  (\ref{9}) is known as the
Adler-Bell-Jackiw anomaly   \cite{5}-\cite{8}.

To better understand the origin of an anomaly, we examine the same problem in the
momentum space. In QED, the matrix element of the transition  of axial current with a
momentum  $q$ into two real or virtual photons with momenta $p$ and $p'$ is described by
the diagrams in Fig.1. The matrix element is
 \be T_{\mu\alpha\beta}(p,p^{\prime}) =
\Gamma_{\mu\alpha\beta}(p,p^{\prime})
+\Gamma_{\mu\beta\alpha}(p^{\prime},p),\label{11}\ee

\begin{figure}[tb]
\hspace{44mm} \epsfig{file=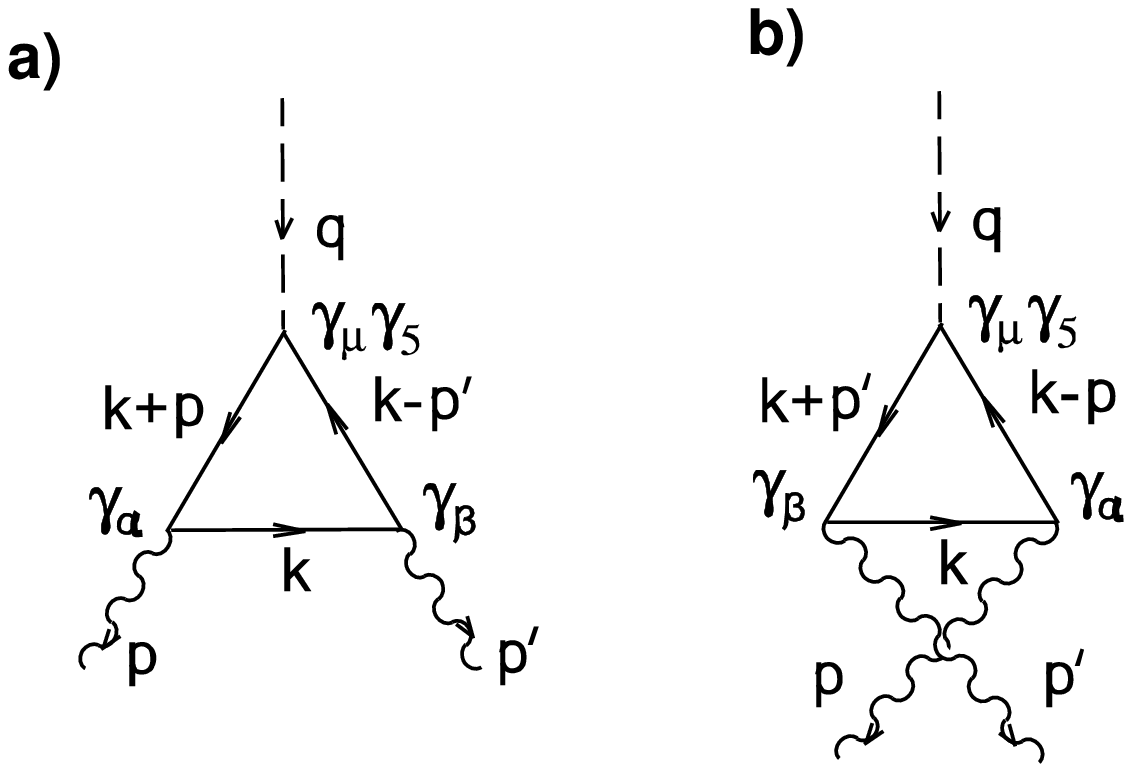, width=68mm}

\vspace{5mm}

{\bf Figure 1.} The Feynman diagrams describing the transition of an axial current with a
momentum $q$ into two real or virtual photons with momenta $p$ and $p'$. $q=p+p'$: (a)
the direct diagram, and (b) the crossing diagram.

\end{figure}
\be \Gamma_{\mu\alpha\beta}(p,p^{\prime}) = -e^2 \int \frac{d^4 k}{(2\pi)^4} Tr \biggl [
\gamma_{\mu}\gamma_5 (\not\!k +\not\!p -m)^{-1}\gamma_{\alpha}(\not\!k -m)^{-1} \gamma_{
\beta}(\not\!k-\not\!p^{\prime} -m)^{-1}\biggr ].\label{12}\ee  Integral (\ref{12})
linearly diverges. In a linearly divergent integral, the important terms are the surface
terms, which emerge as a result of integrating over an infinitely remote surface in the
momentum space. (This becomes especially clear when the vectors   $q,p$, and $p'$ are
space-like and the integration contour over   $k_0$ can be turned to the imaginary axis,
$k_0 \to ik_4$, such  that integration over $k$ is carried out in Euclidean space.) The
result of calculations depends on the way $k$ is chosen: we can displace $k$ by an
arbitrary constant vector  $a_{\lambda}$, i.e.,  $k_{\lambda} \to k_{\lambda}
+a_{\lambda}$. Amplitude (\ref{11}) must satisfy  the conditions needed for the
vector-current conservation: $p_{\alpha} T_{\mu\alpha\beta}(p,p')=0,~p'_{\beta}
T_{\mu\alpha\beta}(p,p')=0.$

We try to choose the vector $a_{\lambda}$ such that the conditions for both axial- and
vector-current conservation are satisfied. We parameterize $a_{\lambda}$ as
$a_{\lambda}=(a+b) p_{\lambda} + bp'_{\lambda}$. The result of calculations shows that
both conditions cannot be satisfied simultaneously: the vector-current conservation can
be achieved at $a=-2$, while the axial-current conservation requires that $a=0$
\cite{8,9}. Vector-current conservation is the necessary  condition for the existence of
QED. Hence, we select $a=-2$. The divergence of the axial current is
\be q_{\mu} T_{\mu\alpha\beta} (p,p^{\prime}) = \biggl [2 m G(p,p^{\prime})
-\frac{e^2}{2\pi^2}\biggr ] \varepsilon_{\alpha\beta\lambda\sigma} p_{\lambda}
p^{\prime}_{\sigma}.\label{13}\ee Here, we restore the term proportional to the electron
mass and define $G(p,p')$ as
 \be \langle p,\varepsilon_{\alpha};p^{\prime},
\varepsilon^{\prime}_{\beta} \mid \bar{\psi} \gamma_5 \psi\mid 0 \rangle =
G(p,p^{\prime}) \varepsilon_{\alpha\beta\lambda\sigma}
p_{\lambda}p^{\prime}_{\sigma},\label{14}\ee with $\varepsilon_{\alpha}$ and
$\varepsilon^{\prime}_{\beta}$ being the polarizations of the two photons. The fact that
the axial current is not conserved, stated in Eqn (\ref{13}), is equivalent to Eqn
(\ref{9}). Our discussion of the axial anomaly in QED was limited to terms of the order
$e^2$. Adler and Bardeen have proved (see Refs \cite{5,6,10}) that the radiative
corrections caused by the photon lines connecting different points inside the triangle
diagrams in Fig.1 do not alter the anomaly equation. However, the diagram in Fig.2 yields
a nonvanishing, albeit small correction of the order  $e^6$ to this condition \cite{11}.

\begin{figure}[tb]
\hspace{65mm} \epsfig{file=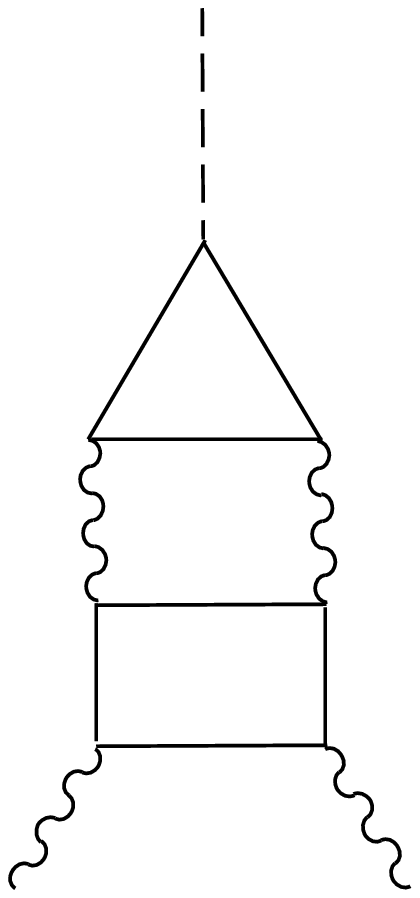, width=25mm}

\vspace{7mm}

\hspace{1.5cm}{\bf Figure 2.} The $e^6$ correction to Adler-Bell-Jackiw anomaly in QED.

\end{figure}
\vspace{7mm}

{\bf \large 3. The axial anomaly and its relation to the structure }

\vspace{2mm}

\hspace{7mm}{\bf \large of the vacuum in quantum chromodynamics}

\bigskip

\noindent In QCD with massless quarks, the axial anomaly is described by a formula
similar to (\ref{9}):
\be
\partial_{\mu} j^a_{\mu 5} = \frac{e^2}{8\pi^2} e^2_q N_c F_{\mu\nu} \tilde{F}_{\mu \nu}.
\label{15}\ee Here, $N_c=3$ is the number of colors and $e_q$ is the quark charge. (We
wrote Eqn  (\ref{15}) for one massless quark.) There is  also another anomaly in QCD,
where the external fields are gluonic rather than electromagnetic:
\be
\partial_{\mu} j_{\mu 5} =\frac{\alpha_s N_c}{4\pi} G^n_{\mu\nu}
\tilde{G}^n_{\mu\nu},\label{16}\ee where $G^n_{\mu\nu}$ is the gluonic field strength and
$\tilde{G}^n_{\mu\nu}$ is its   dual. Equation (\ref{16}) can be considered as  an
operator equation, and the fields $G^n_{\mu\nu}$ and $\tilde{G}^n_{\mu\nu}$ represent the
fields of virtual gluons. In the same way as in QED, perturbative corrections to
(\ref{16}) begin at $\alpha^3_s$ and are described by a diagram similar to the one shown
in Fig.2. In QCD, however, the coupling constant is large, with the result  that the
contribution provided by this diagram  is not small; the contribution of diagrams
obtained from the one in Fig.2 by attaching additional quark and gluon loops are not
small either. Obviously, in QCD, the octet axial current
 \be j^i_{\mu 5} =\sum_q
\bar{\psi}_q \gamma_{\mu}\gamma_5 (\lambda^i/2)\psi_q,~~~i = 1, ...8\label{17}\ee is
conserved in the absence of an electromagnetic field. (Here, $\lambda_i$ is the Gell-Mann
matrix, and sumation is over the flavors of the light quarks, $q=u,d,s$.) The singlet
axial current
 \be j^{(0)}_{\mu 5} = \sum_q \bar{\psi}_q \gamma_{\mu}\gamma_5
\psi_q,\label{18}\ee contains the anomaly
 \be
\partial_{\mu}j^{(0)}_{\mu 5} = 3\frac{\alpha_s N_c}{4\pi}
G^n_{\mu\nu} \tilde{G}^n_{\mu\nu}.\label{19}\ee In view of the spontaneous breaking of
chiral symmetry, the pseudoscalar mesons belonging to the octet ($\pi,K,\eta)$ are
massless (in the $m_q\to 0$ approximation), while the $SU(3)$ singlet $\eta'$ is massive.
In this way, the presence of an anomaly solves what is known as the $U(1)$ problem
\cite{12}.

I now discuss the important assertion that exists in QCD and relates the structure of an
anomaly to the structure of the  vacuum in this theory. Because we deal with the
existence of degenerate vacua and tunnel (underbarrier) transitions between them, it is
convenient (just as in quantum mechanics) to introduce imaginary time by setting $t=x_0 =
-ix_4$, we thus operate in the Euclidean space, where $x^2=x^2_1 + x^2_2 + x^2_3 +
x^2_4$. In the Euclidean space, the action integral
 \be S=\frac{1}{4} \int d^4 x
G^2_{\mu\nu}\label{20}\ee is positive. (We temporarily ignore the quark contribution.)
The transition amplitudes are determined by the matrix elements of  $\exp(-S)$. A theorem
first proved by Belavin, Polyakov, Schwartz,  and Tyupkin  \cite{13}states that
 \be
\frac{\alpha_s}{8\pi} \int d^4 x G^n_{\mu\nu} \tilde{G}^n_{\mu\nu} = n,\label{21}\ee
where $n$ is an integer known as the winding number. Here, I do not prove this theorem in
detail; instead, I mention its main points. The integrand in (\ref{21}) can be written as
the total derivative
\be G^n_{\mu\nu} \tilde{G}^n_{\mu\nu}=\partial_{\mu} K_{\mu},\label{22}\ee
\be K_{\mu} =\varepsilon_{\mu\nu\gamma\delta} (A^n_{\nu} G^n_{\gamma\delta} - \frac{1}{3}
f^{nmp} A^n_{\nu} A^m_{\gamma} A^p_{\delta}).\label{23}\ee When $x^2$ is large,
$G_{\mu\nu}(x)$ decreases faster than  $1/x^2$ (i.e., there is no physical field), and
$A_{\mu}^n$ is a pure-gauge field. Then, we can drop the first term in the right-hand
side of (\ref{23}) and keep only the second term in the general expression for the gauge
transformation for $A^n_{\mu}$
\be A'_{\mu} = U^{-1} A_{\mu} U + i U^{-1}
\partial_{\mu} U\label{24}\ee
(here, $U$ is a unitary, unimodular matrix, $U^+=U$, $\mid U \mid =1$.) We suppose that
the field $A^n_{\mu}(n=1,2,3)$ belongs to the subgroup  $SU(2)$ of the color group
$SU(3)$. This subgroup plays a special role in the $SU(3)$ group because it is isomorphic
to the spatial rotation group  $O(3)$. At this point, it is convenient to introduce
matrix notation for the field $A_{\mu}$: \be A_i =\frac{1}{2} g\tau^k A^k_i,~~~ k=1,2,3;~
i=1,2,3.\label{25}\ee Then according to Eqns (\ref{22}) and (\ref{23}), we have
\be \int d^4 x G_{\mu\nu}(x) \tilde{G}_{\mu\nu} (x) = -i\frac{4}{3} \frac{1}{g^2} \int dV
\varepsilon_{ikl}  Tr (A_i A_k A_l).\label{26}\ee Substituting the second term in the
right-hand side of Eqn  (\ref{24}) into (\ref{26}), we see that the integrand in
(\ref{26}) is a total derivative with respect to the spatial coordinates, and therefore
reduces to an integral over an infinitely remote surface. Because $\mid U\mid =1$ on this
surface, the matrix $U$ has the form
\be U =\exp (2\pi n \hat{r}_a \tau^a/2 i),\label{27}\ee where $n$ un an integer, and
$\hat{r}_a$ is a unit radius vector, $\hat{r}_a =r_a/\mid {\bf r}\mid$. The invariance of
$U$ under spatial rotations stems from the fact that each  such rotations is accompanied
by a gauge transformation, a rotation in the   $SU(2)$ group. When the right hand side of
Eqn (\ref{24}) is substituted in ({26}), we see that Eqn (\ref{21}) follows from
(\ref{27}). Theorem (\ref{21}) also follows from general mathematical considerations,
because group $SU(2)$ is mapped onto  $O(3)$; such a map is multivalued and is determined
by the number of times the $O(3)$ group is covered.   We note  that the fields
corresponding to different
 $n$ cannot be transformed into each other by a continuous transformation.
 In the perturbation theory, we always deal with fields corresponding
to $n=0$. The action integral in  (\ref{20}) can be written as:
\be S =\frac{1}{4} \int d^4 x G^n_{\mu\nu} G^n_{\mu\nu} = \frac{1}{4} \int d^4 x
[~G^n_{\mu\nu} \tilde{G}^n_{\mu\nu} +\frac{1}{2} (G^n_{\mu\nu}-
\tilde{G}^n_{\mu\nu})^2~].\label{28}\ee Because the last term (\ref{28}) is positive, the
minimum of the action is achieved with fields satisfying the self-duality condition
\be G^n_{\mu\nu} = \tilde{G}^n_{\mu\nu},\label{29}\ee
\be S_{min} = \frac{1}{4} \int d^4 x ~G^n_{\mu\nu} \tilde{G}^n_{\mu\nu}=
\frac{8\pi^2}{g^2}\mid n \mid =\frac{2\pi}{\alpha_s} \mid n \mid. \label{30}\ee (Negative
$n$ correspond to anti-self-dual fields $G^n_{\mu\nu} =-\tilde{G}^n_{\mu\nu}$). The
solutions of the self-duality equation (for $n=1$), which became known as instantons,
were found in \cite{13}. It follows from (\ref{30}) that in QCD in the Euclidean space,
there exists an infinite number of action minima. In Minkowski space, instantons are
paths of tunnel transition (in the field space) between vacuum characterized by different
winding numbers but having the same energies \cite{14}-\cite{16}. By examining $n(t)$,
which transforms into the winding number as $t\to \pm \infty$, it can be shown that the
instanton solutions correspons to  $n(t\to -\infty)=0$ andи $n(t\to \infty)=1$ and that
the transition amplitude between states \cite{17}
\be \langle \Omega_{n=1} (t\to \infty) /\Omega_{n=0} (t\to -\infty)\rangle =
e^{-2\pi/\alpha_s}.\label{31}\ee

\vspace{7mm}

{\bf \large 4. Structure of the vacuum in quantum chromodynamics}

\bigskip

\noindent  Above, we showed that in QCD, there is an infinite number of vacua with the
same energies, vacua that are characterized by  the values of the winding number $n$. We
let  $\Omega(n)$ denote the wave function of such a vacuum and suppose that the wave
functions are normalized, $\Omega^+(n)\Omega(n)=1$, and form a complete system. The
ambiguity in the wave function resides in the phase factor,
 $\Omega(n) = e^{i\theta_n} \Omega'(n)$. We separate the Euclidean space into two big parts
 and assume that the field strength in the space between these parts is zero and the
 potentials are pure gauge. Then, obviously,
 \be e^{i\theta_{n_1+n_2}}
\Omega(n_1+n_2) =e^{i\theta_{n_1}}\Omega(n_1) e^{i\theta_{n_2}} \Omega(n_2).\label{32}\ee
[Here, we drop the prime on $\Omega'(n)$.] Because
\be \Omega(n_1+n_2)=\Omega(n_1) \Omega(n_2),\label{33}\ee we have the equation
\be \theta_{n_1+n_1} =\theta_{n_1} +\theta_{n_2},\label{34}\ee whose solution is
\be \theta_n = n\theta.\label{35}\ee Thus, the vacuum wave function in QCD is a linear
combination of wave functions with different winding numbers:
 \be \Omega(\theta) =\sum_n e^{i n\theta}
\Omega(n).\label{36}\ee The state $\Omega(\theta)$ is known as the $\theta$-vacuum. The
vacuum state $\Omega(\theta)$ is similar to the Bloch state of an electron in a crystal,
with  $\theta$ acting as momentum. But in contract to a Bloch state, all transitions
between states with different $\theta$ are forbidden for the $\theta$-vacuum. The vacuum
state $\Omega(\theta)$ can be reproduced if the term
\be L_{\theta} =\frac{g^2\theta}{32 \pi^2} G_{\mu\nu} \tilde{G}_{\mu\nu}.\label{37}\ee is
added to the QCD Lagrangian (in Minkowski space). The presence of this term in the
Lagrangian demonstrates that $\theta$ is an observable. Term (\ref{37}) violates the P-
and  CP-invariance. However, so far all attempts to discover the violation of
CP-invariance in strong interactions have failed. The strongest bound on the value of
 $\theta$ has been found in searches of the neutron dipole moment,
  $\theta <  10^{-9}$ \cite{18}.

\newpage


{\bf \large 5. Zero eigenvalues of the Dirac equation for massless quarks as}

 \vspace{2mm}
\hspace{7mm}{\bf \large a consequence of an anomaly. Spontaneous breaking of chiral}

\vspace{2mm}

\hspace{7mm}{\bf \large symmetry in quantum chromodynamics}

\bigskip

\noindent We consider the Dirac equation for massless quarks in QCD in Euclidean space:
 \be -i\gamma_{\mu}
\nabla_{\mu} \psi_k =\lambda_k\psi_k,~~~~\nabla_{\mu} =\partial_{\mu} + ig
\frac{\lambda_n}{2} A^n_{\mu}. \label{38}\ee From anomaly condition (\ref{16}) with
$n=1$, we have:
 \be \int d^4 x Tr \langle 0\mid \partial_{\mu} j_{\mu 5} (x) \mid 0 \rangle =
\frac{g^2}{16\pi^2} \int d^4 x \langle 0\mid G^a_{\mu\nu} \tilde{G}^a_{\mu\nu} \mid 0
\rangle= 2N_c. \label{39}\ee The left-hand side of Eqn  (\ref{39}) can be written as an
operator as  follows:
$$ \int d^4 x Tr \langle 0\mid \partial_{\mu} j_{\mu 5} (x) \mid 0 \rangle = -\int d^4 x
\partial_{\mu} ~Tr \langle 0 \mid i \not\!{\nabla}^{-1} (x,x) \gamma_{\mu} \gamma_5 \mid 0
\rangle =$$ \be -\int d^4 x \nabla_{\mu}~Tr \biggl [ \sum_{k}
\frac{\psi_{k}(x)\psi^+_{k}(x)}{\lambda_{k}} \gamma_{\mu}\gamma_5\biggr ] = -\int d^4 x
~Tr \biggl [ \sum_k \frac{\psi_k(x) \psi^+_k(x)}{\lambda_k} \cdot 2\lambda_k
\gamma_5\biggr ].\label{40}\ee States with nonzero $\lambda_k$ contribute nothing to
(\ref{40}) because each such state  $\psi_k(x)$ corresponds to the state
$\gamma_5\psi_k(x)$ with the eigenvalue --$\lambda_k$, and the two states are orthogonal.
Thus, only the zero modes contribute, and hence we have
 \be 2\int d^4 x Tr [~\gamma_5 \psi_0 (x) \psi^+_0(x)~] =
-2N_c.\label{41}\ee This implies that in the case where $n=1$, in the instaton field, the
zero mode is right-handed: the quark spin is directed  along the quark momentum,
$\gamma_5\psi_0 = -\psi_0$. (Actually, for a quark in the instanton field, only one
right-handed zero mode exists, because spin is correlated with color and the factor $N_c$
in the right-hand side of Eqn  (\ref{41}) disappears.)  At $n=-1$, the resulting equation
differs from   (\ref{41}) only in sign; i.e., a left-handed zero mode exists in an
anti-instanton field. In the  general case, we have the Atiyah-Singer theorem \cite{19},
according to which
 \be n=n_L - n_R,\label{42}\ee where $n_L$ and  $n_R$ are the numbers of left-- and
 right-hand zero modes respectively. It follows from
 (\ref{41}) that in an instanton  field, the zero mode violates the chiral symmetry of the
Lagrangian, i.e., the invariance under the transformations $\psi \to \gamma_5 \psi$. (We
note that in passing from the Euclidean metric to Minkowski space, the function
   $\psi^+$ is replaced by $\bar{\psi}$.) Thus the presence of instantons is an
   indication that a quark condensate exists in the QCD vacuum:
 \be \langle 0 \mid \bar{\psi} \psi \mid 0 \rangle \not= 0,\label{43}\ee
which breaks the chiral symmetry of the Lagrangian. (Unfortunately, it is impossible to
calculate the quark condensate on the basis of the instanton approach, because this
approach is meaningful only when the distances are small, while the condensate forms over
large distances.)

The winding number  $n$ corresponds to the topological current operator
 \be Q_5(x)= \frac{\alpha_s}{8\pi} G^n_{\mu
\nu}(x)\tilde{G}^n_{\mu\nu}(x).\label{44} \ee It was found in \cite{20} that the vacuum
correlator of topological currents
 \be \zeta(q^2) = i\int d^4x e^{iqx} \langle
0\mid T\left \{ Q_5(x),~Q_5(0)\right \}\mid \rangle \label{45}\ee vanishes at $q^2=0$ if
the theory contains at least one massless quark. Later, it was proved in  \cite{21} that
in the limit as  $N_c \to \infty$ the relation
\be \zeta(0) = \langle 0\mid \bar{q}q\mid 0\rangle \Biggl (
\sum^{N_f}_i~\frac{1}{m_i}\Biggr )^{-1}\label{46} \ee holds. In the cases of two and
three massless quarks, the validity of Eqn (\ref{46}) was proved in Ref.\cite{22}, where
the limit  $N_c \to \infty$ was not used. The concept of topological current turned out
to be highly effective in QCD: it has been used to  establish the spin composition of the
proton  \cite{23}, to establish a relation between the spin structure functions for large
and small установить связь между спиновыми структурными функциями при больших и малых
$Q^2$ \cite{24},\cite{25}, and to determine the axial coupling constants for the nucleon
 \cite{26}.

\vspace{7mm}

{\bf \large 6. The sum rule for the axial anomaly in quantum }

\vspace{2mm}

\hspace{7mm}{\bf \large chromodynamics}

\bigskip

\noindent We consider the  general representation of the transition amplitude of the
axial current into two photons with momenta  $p$ and $p'$ in terms of the structure
functions (form factors) without kinematic singularities,  $T_{\mu\alpha\beta} (p,p')$
\cite{27}. We limit ourselves to the case where  $p^2=p^{\prime 2}$. Then \cite{28,29}
$$ T_{\mu\alpha\beta}(p,p^{\prime}) = F_1(q^2,p^2)
q_{\mu}\varepsilon_{\alpha\beta\rho\sigma} p_{\rho}p^{\prime}_{\sigma} - $$ \be
-\frac{1}{2} F_2(q^2,p^2)\biggl
[\varepsilon_{\mu\alpha\beta\sigma}(p-p^{\prime})_{\sigma} -\frac{p_{\alpha}}{p^2}
\varepsilon_{\mu\beta\rho\sigma} p_{\rho}p^{\prime}_{\sigma} +
\frac{p^{\prime}_{\beta}}{p^2} \varepsilon_{\mu\alpha\rho\sigma}
p_{\rho}p^{\prime}_{\sigma}\biggr ].\label{47}\ee The anomaly condition in QCD reduces to
 \be F_2(q^2,p^2) + q^2F_1(q^2,p^2) = 2\sum_q m_q
G(q^2,p^2) - \frac{e^2}{2\pi^2} \sum_q e^2_q N_c. \label{48} \ee Because
$T_{\mu\alpha\beta} (p,p')$ has no singularity at  $p^2=0$, we have $F_2(q^2,0)=0$. The
functions  $F_1(q^2,p^2),F_2(q^2,p^2)$, and $G(q^2,p^2)$ can be described by dispersion
relations in $q^2$ with no substractions. Using these relations, we can prove the sum
rule
 \be \int\limits^{\infty}_{4m^2} Im~ F_1(t,p^2)dt = \frac{e^2}{2\pi^2} \sum e^2_q
N_c, \label{49}\ee where $m^2$ is the smallest of quark masses. The  sum rule in
(\ref{49}) was proved in  \cite{30} for $p^2 <0, m=0$, in \cite{28} for  $p^2=p^{\prime
2}$ and in \cite{31} in the general cases where  $p^2\not=p^{\prime 2}$. We note that
 (\ref{49}) also holds for massive quarks. We consider the most interesting case where the
 axial current is the third component of the isovector current:
 \be j^{(3)}_{\mu 5} =\bar{u} \gamma_{\mu}\gamma_5 u -
\bar{d}\gamma_{\mu}\gamma_5d.\label{50}\ee We ignore the masses of the $u$- and
$d$-quarks and assume that  $p^2=p^{\prime 2}=0$. Combining (\ref{47}) and (\ref{48}), we
obtain
 \be T_{\mu\alpha\beta} (p,p^{\prime}) =-\frac{2\alpha}{\pi} N_c
\frac{q_{\mu}}{q^2} (e^2_u -e^2_d) \varepsilon_{\alpha \beta\lambda\sigma}
p_{\lambda}p^{\prime}_{\sigma}.\label{51}\ee As follows from  (\ref{51}) the transition
of the isovector axial current into two photons occurs through an intermediate massless
state. Such a state (in the limit $m_u,m_d\to 0$) is the $\pi^0$-meson (Fig.3).
\begin{figure}[tb]
\hspace{60mm} \epsfig{file=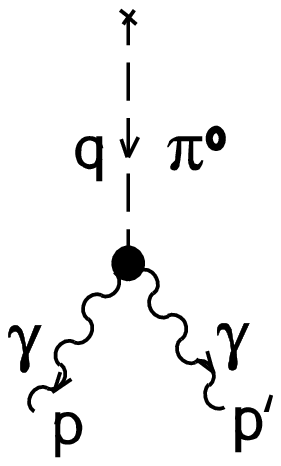, width=25mm}

 {\bf Figure 3.} The diagram describing the
transition of of isovector axial current (denoted by cross) into two photons.

\end{figure}
Combining the fact that  $\langle 0\mid j^{(3)}_{\mu 5} \mid \pi^0 \rangle = \sqrt{2} i
f_{\pi} q_{\mu}$ and the anomaly condition, we can find the matrix element of the $\pi^0
\to 2\gamma$ decay,
 \be M(\pi^0 \to 2\gamma) =A\varepsilon_{\alpha \beta \lambda \sigma}
\varepsilon_{1\alpha}\varepsilon_{2\beta}p_{1\lambda}p_{2\sigma},\label{52}\ee determine
the constant $A$, and calculate the width of   $\pi^0 \to 2\gamma$ as
 \be \Gamma(\pi^0 \to
2\gamma) =\frac{\alpha^2}{32 \pi^3} \frac{m^3_{\pi}}{f^2_{\pi}}.\label{53}\ee The result
was first obtained in \cite{32}. Under the assumption that $f_{\pi_0} = f_{\pi^+} =
130.7$ MeV, we obtain    $\Gamma (\pi^0 \to 2\gamma)_{\mbox{theory}} = 7.73$ eV from
(\ref{53}). It is difficult to estimate the accuracy of the prediction, but apparently it
varies between 5 and 10\%. The experimental value of this quantity averaged over all
existing measurements (data of the year 2006) is  $\Gamma(\pi^0 \to 2\gamma) = 7.8\pm
0.6$ eV \cite{33}. To achieve better accuracy for the theoretical prediction, we must (a)
insert  $f_{\pi^0}$ instead of  $f_{\pi^+}$ in (\ref{53}), and (b) allow the contribution
of excited states (in addition to  $\pi^0$) to the sum rule (\ref{49}) for the isovector
current at $p^2=0$. This program was implemented in Ref.\cite{34}, where it was shown
that the difference  $\Delta f_{\pi} = f_{\pi^0} - f_{\pi^+}$ is small: $\Delta
f_{\pi}/f_{\pi} \approx -1.0 \cdot 10^{-3}$. Among the excited states, only the
$\eta$-meson contributes significantly. Its contribution is determined by the value of
the  $\pi^0-\eta$ mixing angle \cite{35,36} and the width  $\Gamma(\eta\to 2\gamma) =
510$ eV \cite{33}. It was found nn Ref. \cite{34} that $\Gamma(\pi^0 \to
2\gamma)_{\mbox{theory}} = 7.93 \pm 1.5\%$. The most recent measurements in \cite{37}
yield $\Gamma(\pi^0 \to 2\gamma)_{\mbox{exp}} = 7.93 \pm 2\% \pm 2.1\%$ i.e., the
experimental data are in extremely good agreement with the theoretical predictions.

It would seem that Eqn  (\ref{51}) suggests that the existence of a massless (in the
limit of massless  $u$- and $d$-quarks) Goldstone $\pi^0$-meson is a consequence of the
axial anomaly  described by the triangle diagrams in Fig.1. This is not the case,
however. A direct calculation of  $Im ~F_1(q^2,p^2)$ (it is to this function that  the
intermediate $\pi^0$-meson contributes) for $p^2\not= 0$ shows \cite{9,28} that in this
case,  $Im~F_1(q^2,p^2)$ is a regular function of $q$ that tends to a constant as  $q^2
\to 0$ and has no singularities of the $\delta(q^2)$ type, in contrast to the case of
$p^2=0$ described above. Thus, the amplitude  $T_{\mu\alpha\beta}(p,p')$ corresponding to
the transition of the axial current to two virtual photons and calculated according to
the diagrams in Fig.1 has no pole in  $q^2$ at $q^2=0$. On the other hand, basing our
reasoning on a chiral effective theory (e.g., see Ref. \cite{38a}), we can state that the
transition amplitude of the axial current to two virtual photons  must contain the
contribution provided by the  intermediate massless  $\pi^0$-meson (see Fig.3). As shown
in \cite{6}, the introduction of gluon lines into the diagrams in Fig.1 does not change
the expression for the anomaly. (Actually, this was shown in  \cite{6} to be true for
QCD, but there is no difference between QCD and QED in this aspect.) Thus, from examining
the case  where   $p^2\not= 0$, we conclude that the appearance of a massless
$\pi^0$-meson in the dispersion representation of the AVV form factor is not caused by an
anomaly. The presence of massless Goldstone mesons ($\pi,K,\eta)$ stems from the
spontaneous breaking of chiral symmetry in the QCD vacuum. That there is a singularity at
 $q^2=0$ in the amplitude  $T_{\mu\alpha\beta}(p,p')$ when $p^2=0$ is sometimes interpreted
 as the double nature of the anomaly, the ultraviolet and the  infrared (e.g., see Ref.
   \cite{3}). I believe that in view of the absence of such a singularity when
$p^2\not= 0$, this interpretation is faulty: the nature of an anomaly in QED and QCD
stems from ultarviolet divergences, the singularity in the amplitudes at small distances.
(In this respect, QED and QCD differ dramatically from the two-dimensional Schwinger
model, in which the origin of an anomaly is truly double (see Ref. \cite{3}).

For the eight component of the octet current, the transition amplitude of the axial
current to two real  photons, $F_1(q^2,0)$, has a pole at  $q^2=0$ if  $m_u=m_d=m_s=0$.
It is only natural to associate this pole  with the $\eta$-meson. However, a relation for
$\Gamma(\eta \to 2\gamma)$ similar to (\ref{53}) differs dramatically from the
experimental result. A possible explanation of such a discrepancy is the strong
nonperturbative interaction of the type of instantons in a  pseudoscalar channel mixing
$\eta$ and $\eta'$-mesons \cite{38}. In the case of a singlet axial current, the
amplitude  $j_{\mu 5}^{(0)} \to 2\gamma$ contains diagrams of the type shown in Fig.2
(with virtual gluons instead of photons), their extensions, and nonperturbative
contributions. Hence , we cannot expect reliable predictions concerning the width of
$\eta' \to 2\gamma$ based on  anomalies.

$'$t Hooft \cite{40} hypothesized that the singularities of the amplitudes calculated in
QCD on the quark -- gluon basis should reproduce themselves in calculations on the hadron
basis. Obviously, this is true if both perturbative and nonperturbative interactions are
taken into account. However, as a rule we know nothing about the nonperturbative
interactions. In the cases discussed above (expect for the decay of $\pi^0$ into two real
photons), $'$t Hooft's hypothesis does not hold \cite{9}.

\vspace{7mm}

\centerline{\bf \large 7. Conclusion}

\bigskip

 1. An anomaly is an important and necessary element of quantum field theory.

2. An anomaly emerges because the amplitudes of quantum field theory contain ultraviolet
singularities, in view of which it is necessary to augment the Lagrangian by
renormalization conditions.

3. An anomaly in QCD is connected with appearance  of a new quantum number, the winding
number.

4.The vacuum in QCD is a linear combination of an infinite number of vacua with different
winding numbers.

5. Transitions between vacua with different winding numbers are tunnel transitions
occurring along classical paths  in the field space, self-dual solutions of QCD
equations, or instantons.

6. The axial anomaly in QCD results  in the appearance of zero modes in the Dirac
equations for light quarks and points to the existence of spontaneous breaking of chiral
symmetry in the QCD vacuum, the existence of a quark condensate.

7. The axial anomaly predicts the width of the $\pi^0 \to 2 \gamma$ decay with  a high
accuracy ($\sim 2\%$), a result corroborated by experiments.

This work was supported in part by   RFBR grant 06-02-16905a and the funds from EC to the
project ``Study of Strongly Interacting Matter'' under contract No. R113-CT-2004-506078.


\vspace{1cm}
\begin{thebibliography}{99}
\bibitem{1} Treiman S B, Jackiw R, Zumino B and
Witten E {\it Current Algebra and Anomalies} (Princeton Series in Physics, Princeton
University Press, 1985)
\bibitem{2} Collins J {\it Renormalization} (Cambridge
University Press, Cambridge, 1984)
\bibitem{3} Shifman M A  Anomalies in Gauge Theories
{\it Phys. Rep.}  {\bf 209} 343 (1991)
\bibitem{4} Peskin M E and Schroeder D V  {\it An Introduction
to Quantum Field Theories} (Addison-Wesley Publ. Company, 1995)
\bibitem{5} Adler S L {\it Phys. Rev.} {\bf 177} 2426 (1969)
\bibitem{6} Adler S L and Bardeen W {\it Phys. Rev.}
{\bf 182} 1517 (1969)
\bibitem{7} Bell J and Jackiw R {\it Nuovo Cim.} {\bf A51} 47
(1969)
\bibitem{8} Jackiw R {\it Field Theoretical Investigations in
Current Algebra}  (Ref.1, p.p.81-210)
\bibitem{9} Ioffe B L {\it Intern. J. of Mod. Phys.} {\bf 21} 6249 (2006)
\bibitem{10} Adler S L  Anomalies to all orders, in: {\it Fifty years
of Yang-Mills theory} (Ed. by G. t'Hooft) (World Scientific, 2005) p.p.187-228;
hep-th/0405040
\bibitem{11} Anselm A A and Iogansen A {\it JETP Lett.}
{\bf 49} 214 (1989)
\bibitem{12} Weinberg S {\it Phys. Rev.} {\bf D11} 3583 (1975)
\bibitem{13} Belavin A A, Polyakov A M, Schwarz A S and Tyupkin Yu S
{\it Phys. Lett.} {\bf 59B} 85 (1975)
\bibitem{14} Gribov V N, unpublished
\bibitem{15} Jackiw R and Rebbi C {\it Phys. Rev. Lett.} {\bf 37} 172 (1976)
\bibitem{16} Callan C, Dashen R and Gross D {\it Phys. Lett.} {\bf 63B} 334 (1976)
\bibitem{17} Bitar K M and Chang S -J {\it Phys. Rev.} {\bf D17} 486 (1978)
\bibitem{18} Alfarev  I S et al. {\it Phys. Lett.} {\bf B276} (1992) 242
\bibitem{19} Atiyah M and Singer I {\it Ann. Math.} {\bf 87} 484 (1968); {\bf 93}
119 (1971)
\bibitem{20} Crewther R J {\it Phys. Lett.} {\bf 70B} 349 (1977)
\bibitem{21}  Di Veccia P and  Veneziano G {\it Nucl. Phys.} {\bf B171}
253 (1980)
\bibitem{22} Ioffe B L {\it Yad. Fiz.} {\bf 62} 2226 (1999) [{\it Phys. At. Nucl.} {\bf 62}
2052 (1999)]
\bibitem{23} Ioffe B L and Oganesian A G {\it Phys. Rev.} {\bf D57} R6590 (1998)
\bibitem{24} Burkert V D and Ioffe B L {\it Phys. Lett} {\bf B296} 223 (1992)
\bibitem{25} Burkert V D and Ioffe B L {\it Zh. Eksp. Teor. Fiz.} {\bf 105} 1153 (1994)
[{\it JETP} {\bf 78} 619 (1994)]
\bibitem{26} Ioffe B L {\it  Survey in High Energy Physics} {\bf 14} 89 (1999),
hep-ph/9804238
\bibitem{27} Eletsky V L, Ioffe B L and Kogan Ya I
{\it Phys. Lett.} {\bf B122} 423 (1983)
\bibitem{28}Ho$\check{\mbox{r}}$ej$\check{\mbox{s}}$i J
{\it Phys. Rev.} {\bf D32} 1029 (1985)
\bibitem{29} Bass S D, Ioffe B L, Nikolaev N N and Thomas
A V {\it J. Moscow Phys. Soc.} {\bf 1} 317 (1991)
\bibitem{30} Frishman Y, Schwimmer A, Banks T and
Yankelowicz S {\it Nucl. Phys.} {\bf B117} 157 (1981)
\bibitem{31} Veretin O L and Teryaev O V {\it Yad. Fiz.}
{\bf 58} 2266 (1995) [{\it Phys. At. Nucl.} {\bf 58} 2150 (1995)]
\bibitem{32} Dolgov  A D  and Zakharov V I  {\it Nucl.
Phys.} {\bf B27} 525 (1971)
\bibitem{33} Yao W-M et al. (Particle Data Group) ``Review of Particle Physics''
{\it J. Phys. G:  Nucl. Part. Phys.} {\bf 33} 1 (2006)
\bibitem{34} Ioffe  B L and Oganesian  A G {\it Phys. Lett.} {\bf B647} 389 (2007)
\bibitem{35} Ioffe B L {\it Yad. Fiz.} {\bf 29} 1611 (1979) [{\it Sov. J. Nucl. Phys.}
{\bf 20} 827 (1979)]
\bibitem{36} Gross D J, Treiman S B and Wilczek F {\it Phys. Rev.} {\bf D19} 2188 (1979)
\bibitem{37} De Jager K et al. (PriMex Collab.) arXiv: 0801.4520
\bibitem{38a} Ioffe B L {\it Usp. Fiz. Nauk} {\bf 171} 1273 (2001) [{\it Phys. Usp.} {\bf
44} 1211 (2001)]
\bibitem{38} Geshkenbein B V and Ioffe B L {\it Nucl. Phys.} {\bf B166} 340 (1980)
\bibitem{40} $^{\prime}$t Hooft G in: {\it Recent Developments in Gauge
Theories} (Eds. G.t'Hooft  et al.)  Plenum Press, N.Y., 1980, p.241

\end{thebibliography}
\end{document}